\documentclass{article}
\usepackage{spconf,amsmath,graphicx}
\usepackage{times}
\usepackage{epsfig}
\usepackage{amssymb}
\usepackage{booktabs}
\usepackage{caption}
\usepackage{multirow}
\usepackage{subcaption}
\usepackage{caption}
\usepackage{tikz}
\usepackage{mathtools}
\usepackage{stackengine,tikz,graphicx}
\usetikzlibrary{calc}

\usepackage{enumitem}
\setlist{nosep, leftmargin=14pt}

\usepackage{mwe} 

\title{Enhanced Sharp-GAN For Histopathology Image Synthesis}
\name{Sujata Butte$^\dagger$, Haotian Wang$^\dagger$\thanks{$^\dagger$ equal contribution}, Aleksandar Vakanski, Min Xian$^*$ \thanks{$^*$ corresponding author, mxian@uidaho.edu.}}
\address{University of Idaho, Idaho, USA}
\begin{document}
%
\maketitle
\begin{abstract}
Histopathology image synthesis aims to address the data shortage issue in training deep learning approaches for accurate cancer detection. However, existing methods struggle to produce realistic images that have accurate nuclei boundaries and less artifacts, which limits the application in downstream tasks. To address the challenges, we propose a novel approach that enhances the quality of synthetic images by using nuclei topology and contour regularization. The proposed approach uses the skeleton map of nuclei to integrate nuclei topology and separate touching nuclei. In the loss function, we propose two new contour regularization terms that enhance the contrast between contour and non-contour pixels and increase the similarity between contour pixels. We evaluate the proposed approach on the two datasets using image quality metrics and a downstream task (nuclei segmentation). The proposed approach outperforms Sharp-GAN in all four image quality metrics on two datasets. By integrating 6k synthetic images from the proposed approach into training, a nuclei segmentation model achieves the state-of-the-art segmentation performance on TNBC dataset and its detection quality (DQ), segmentation quality (SQ), panoptic quality (PQ), and aggregated Jaccard index (AJI) is 0.855, 0.863, 0.691, and 0.683, respectively.

\end{abstract}
\section{Introduction}
\label{sec:intro}

Histopathology nuclei analysis provides the golden standard for cancer detection and cancer grading \cite{he2012histology}. Automatic nuclei segmentation extracts crucial morphological information to diagnose cancer. Recent nuclei segmentation methods \cite{kumar2017dataset, naylor2018segmentation,graham2019hover, wang2020bending} achieved state-of-the-art performance in nuclei segmentation by using deep learning approaches. However, these methods heavily relied on large, labeled, and high-quality datasets; and existing datasets \cite{kumar2017dataset, naylor2018segmentation, graham2019hover, vu2019methods} in nuclei analysis are small because of the expensive and time-consuming data preparation process.

Recent generative adversarial networks (GANs) have demonstrated that high-quality synthetic images could improve the segmentation performance \cite{hou2019robust, butte2022sharp}. Hou et al. \cite{hou2019robust} proposed a GAN that learned the simulated texture information from real images and synthesized realistic histopathology images from synthetic masks. Sharp-GAN et al. \cite{butte2022sharp} proposed a GAN-based loss function to maximize the color contrast between background stroma and nuclei for generating accurate nuclei. However, existing methods still cannot generate accurate nuclei boundaries (Fig. 1). Generating realistic clustered and overlapped nuclei are important because they are mostly cancerous nuclei in real images. 

\begin{figure}
\begin{center}
\rotatebox{90}{\hspace{0.5cm}(a)}
  \begin{subfigure}[b]{0.21\linewidth}
    \includegraphics[width=17mm, height=14mm]{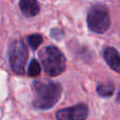}
  \end{subfigure}
  \begin{subfigure}[b]{0.21\linewidth}
    \includegraphics[width=17mm, height=14mm]{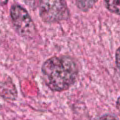}
  \end{subfigure}
  \begin{subfigure}[b]{0.21\linewidth}
    \includegraphics[width=17mm, height=14mm]{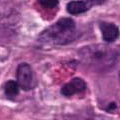}
  \end{subfigure}
  \begin{subfigure}[b]{0.21\linewidth}
    \includegraphics[width=17mm, height=14mm]{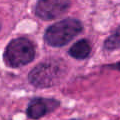}
  \end{subfigure}
    \\
  \rotatebox{90}{\hspace{0.5cm}(b)}
   \begin{subfigure}[b]{0.21\linewidth}
    \includegraphics[width=17mm, height=14mm]{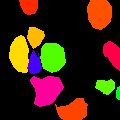}
  \end{subfigure}
  \begin{subfigure}[b]{0.21\linewidth}
    \includegraphics[width=17mm, height=14mm]{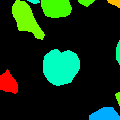}
  \end{subfigure}
  \begin{subfigure}[b]{0.21\linewidth}
    \includegraphics[width=17mm, height=14mm]{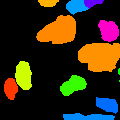}
  \end{subfigure}
  \begin{subfigure}[b]{0.21\linewidth}
    \includegraphics[width=17mm, height=14mm]{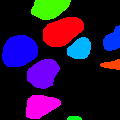}
  \end{subfigure}
  \\
  \rotatebox{90}{\hspace{0.5cm}(c)}
  \begin{subfigure}[b]{0.21\linewidth}  
    \includegraphics[width=17mm, height=14mm]{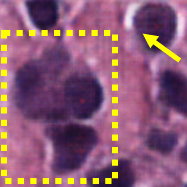}
  \end{subfigure}  
  \begin{subfigure}[b]{0.21\linewidth}
     \includegraphics[width=17mm, height=14mm]{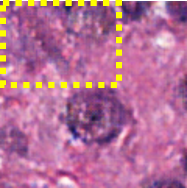}     
  \end{subfigure}  
  \begin{subfigure}[b]{0.21\linewidth}
    \includegraphics[width=17mm, height=14mm]{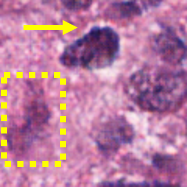}
  \end{subfigure}  
  \begin{subfigure}[b]{0.21\linewidth} 
    \includegraphics[width=17mm, height=14mm]{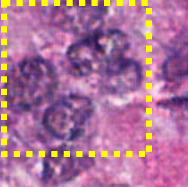}
  \end{subfigure} 
  \\
  \rotatebox{90}{\hspace{0.5cm}(d)}
  \begin{subfigure}[b]{0.21\linewidth} 
    \includegraphics[width=17mm, height=14mm]{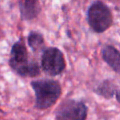}
  \end{subfigure}
  \begin{subfigure}[b]{0.21\linewidth}
    \includegraphics[width=17mm, height=14mm]{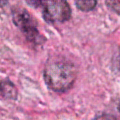}
  \end{subfigure}    
  \begin{subfigure}[b]{0.21\linewidth} 
    \includegraphics[width=17mm, height=14mm]{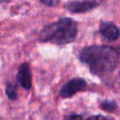}
  \end{subfigure}
  \begin{subfigure}[b]{0.21\linewidth}   
    \includegraphics[width=17mm, height=14mm]{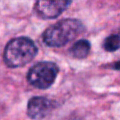}
  \end{subfigure}

\end{center}
   \vspace*{-5mm}
   \caption{(a) Real images. (b) Nuclei masks. (c) Synthesized image patches from Sharp-GAN \cite{butte2022sharp}. Yellow rectangles highlight the clustered regions and arrows point to image artifacts. (d) our synthesized image patches. }
\label{fig:long}
\label{fig:onecol}
\end{figure}
 
\begin{figure*}[t]
    \centering    
    \includegraphics[width=0.9\textwidth]{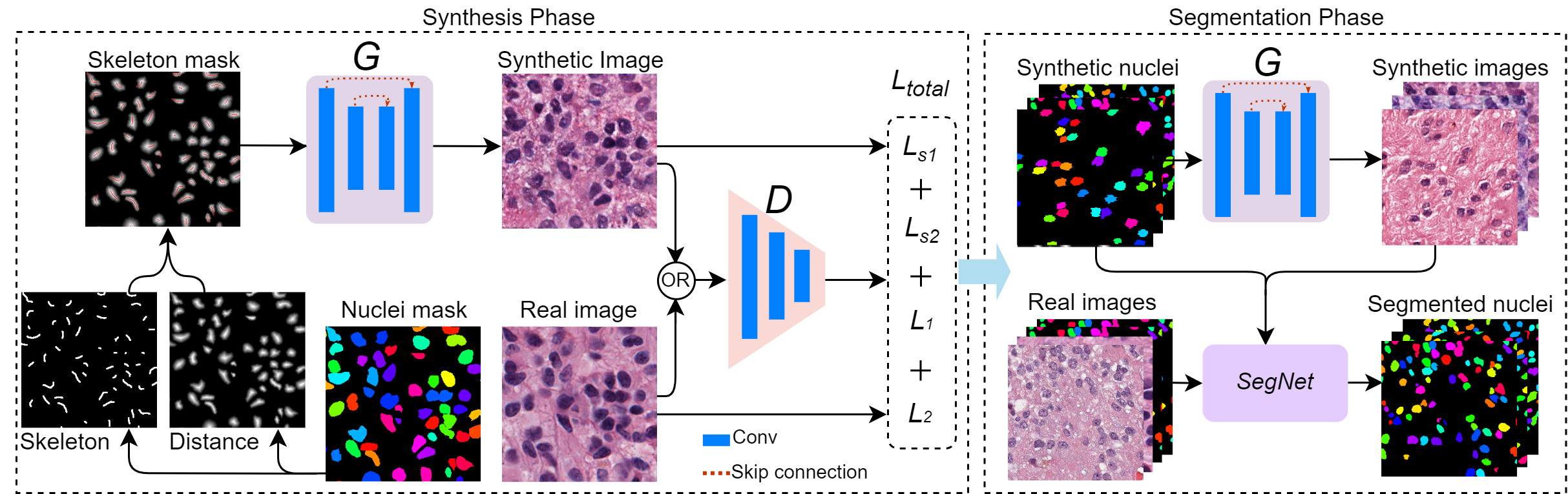}
        \vspace*{-3mm}
    \caption{Overview of proposed model Enhanced Sharp-GAN}
    \label{fig:Fig2}
\end{figure*}

To address this issue, we propose the enhanced sharpness loss regularized GAN to generate realistic histopathology images with clear nuclei contours. One of the shortcomings of using binary masks as the input of a generator is that no clear boundaries are provided for clustered nuclei, and it is difficult for the generator to restore the boundaries. Therefore, we replace binary masks with skeleton map of nuclei which embeds nuclei topology and contours. Furthermore, two novel contour regularization terms are proposed to enhance the contrast between contour and non-contour pixels and to increase the smoothness of contour pixels.. 

The proposed method is discussed in detail in Section 2. Section 3 discusses the experimental results. and the conclusion in Section 4.

\section{Proposed Method }
\label{sec:format}

The architecture of the proposed network is shown in Fig. 2.  The generator \(G\) is conditioned on a skeleton map of nuclei \(x\) to generate a corresponding histopathology image \(y\). The discriminator \(D\) is trained to discriminate between synthetic and real histopathology images. The proposed skeleton map combines nuclei skeletons and distance maps, which preserve nucleus topology and distinguish densely clustered nuclei. By using the skeleton map, the generator \(G\) will be informed to learn nuclei topology and boundaries. In the loss function, two contour regularization terms ($L_{s1}$ and $L_{s2}$) increase the contrast between contour and non-contour pixels and increase the smoothness between contour pixels. At the segmentation phase, we introduce segmentation as a downstream task, and apply synthetic augmentation strategies using both real and synthetic images generated at the synthesis phase to train SegNet. 

\subsection{Contour regularization}

In Sharp-GAN \cite{butte2022sharp}, the sharpness loss term was proposed to generate clear boundaries of nuclei. It defined high penalties to contour pixels with small contrast to their neighboring pixels. The sharpness loss alleviated the blurry boundary issue of GANs. However, it treated all neighboring pixels equally and attempted to increase their contrast, which also increased the contrast between adjacent contour pixels and led to non-smooth nuclei contour. To address the issues, We propose two new contour regularization terms: $L_{s1}$ and $L_{s2}$. $L_{s1}$ defines loss between neighboring contour pixels and attempts to enhance the smoothness of a contour; and $L_{s2}$ defines loss between contour and non-contour pixels and aims to increase their contrast.  

The proposed total loss is defined by
        
        \begin{equation}\label{1}
        L(G, D) = E_{x,y}[L_1] + E_x[L_2 + L_{s1} + \beta L_{s2}]
        \end{equation}                                                
where
                 \begin{equation}\label{2}
                 L_1 = -log D(x,y)
                 \end{equation}

                \begin{equation}\label{3}
                L_2 = -log(1-D(x, G(x)))
                \end{equation}

In Eqs. (1-3), the generator $G$ minimizes the total loss \(L\) against a discriminator \(D\) that maximizes the total loss; \(L_{s1}\) is the proposed smoothness loss term that increases the similarity between contour pixels; \(L_{s2}\) is the new sharpness loss that defines losses only between contour and non-contour pixels; and the parameter \(\beta\) controls the contribution of \(L_{s2}\). Let $c$  be a set of contour pixels(coordinates), and $y$ be a synthesized image of \(G(x)\). The smoothness loss is given by
\begin{equation}\label{4}
  L_{s1}(c, y) = \frac{1}{n} \sum_{(i,j) \in c} S_1 (i, j),
\end{equation}
and the new sharpness loss is

\begin{figure}
\begin{center}
  \begin{subfigure}[b]{0.24\linewidth}
    \vspace*{-2.5mm}    
    \includegraphics[width=\linewidth]{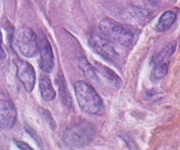}
  \end{subfigure}
  \begin{subfigure}[b]{0.24\linewidth}
    \vspace*{-2.5mm}    
    \includegraphics[width=\linewidth]{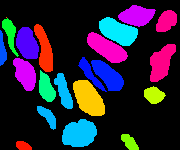}
  \end{subfigure}
  \begin{subfigure}[b]{0.24\linewidth}
  
    \vspace*{-2.5mm}    
    \includegraphics[width=\linewidth]{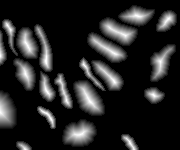}
  \end{subfigure}
  \begin{subfigure}[b]{0.24\linewidth}
    \vspace*{-2.5mm}    
    \includegraphics[width=\linewidth]{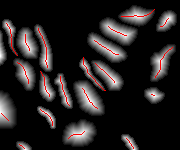}
  \end{subfigure}
\end{center}
    \vspace*{-5mm}
   \caption{From left to right: image patch, nuclei mask, distance mask, and distance mask with skeletons (red).}
\label{fig:long}
\label{fig:onecol}
\end{figure}

\begin{equation}\label{5}
L_{s2}(c, y) = \frac{1}{n} \sum_{(i,j) \in c} S_2(i, j)
\end{equation}
where

\begin{equation}\label{6}
  S_1(i, j) = \sum\limits_{\substack{(p,q)\in N_{i,j} \\  (p,q)\in c}}  \frac{2}{1+exp\frac{-(g_{i,j} - g_{p,q})^2} {\lambda^2}} - 1
\end{equation}

and
\begin{equation}\label{7}
S_2(i, j) = \sum\limits_{\substack{(p,q)\in N_{i,j}\\ (p,q)\notin c}}exp\frac{-(g_{i,j} - g_{p,q})^2}{2\lambda^2} . \frac{1}{dist}
\end{equation}

In above equations, \(n\) is the number of contour pixels of set $c$; and $g$ is the gray scale image of generated image. \(g_{i,j}\) and  \(g_{p,q}\) are the intensities of pixels at position \((i,j)\) and \((p,q)\), respectively. \(N_{i,j} \) is the set of neighboring positions of \((i, j)\) . Eq. (6) is calculated for neighboring pixels on contour. Eq. (7) is calculated on non contour neighbouring pixels. The new sharpness loss gives high penalty for pixels when $(g_{i,j} - g_{p,q})< \lambda$; and it defines small penalty when  $(g_{i,j} - g_{p,q})> \lambda$. We increase the smoothness of contour pixels by adding penalty when pixels are dissimilar. We apply the eight nearest neighboring system to define neighbors.

\subsection{Skeleton map }
The existing approaches \cite{isola2017image, butte2022sharp} use binary image or distance map as input for conditional GAN to generate synthetic images. However, in histopathology images, many nuclei are densely clustered and overlapped. Binary masks cannot show the boundary between two touching nuclei. If we use binary mask as input, the generator may fail to learn accurate boundaries. Though using distance map \cite{butte2022sharp} improved the results, it failed for heavily clustered regions. To address this, we propose the skeleton map. The skeleton map combines nuclei skeletons and distance maps. By using the skeleton map, the generator \(G\) will be informed to learn nuclei topology and boundaries between nuclei.

The distance map is generated by iterating the one-pixel morphological erosion operation starting through nuclei contour. The number of iterations from the contour to its topological skeleton is normalized to form the distance map. The topological skeleton is a set of points having more than one closest point on its contour. Finally. we obtain the skeleton map by calculating the pixel-wise summation of the distance map and skeleton. Fig. 3 demonstrates the process of obtaining skeleton maps. We adopted the nuclei mask generation method in \cite{hou2019robust}. The method randomly generates nucleus-like polygons with variable size and irregularities. The generated polygons can be set to overlap or touch to each other.

\section{EXPERIMENTAL RESULTS}
\label{sec:typestyle}
\subsection{Datasets, metrics and setting}
\textbf{Dataset}. We use three public histopathology image datasets: CPM-15, CPM-17 \cite{vu2019methods}, and TNBC \cite{naylor2018segmentation} in this work. We merge CPM-15 and CPM-17 datasets as one set in our experiments because they are organized by the same team. CPM-15 \& 17 contains 47 histopathology image patches with a total number of 10,475 nuclei. TNBC includes 50 histopathology image patches with a total number of 4,056 nuclei.
   
\textbf{Evaluation metrics}. Three image quality assessment metrics (structural similarity index (SSIM) \cite{wang2004image}, feature similarity index (FSIM) \cite{zhou2017evaluation}, and gradient magnitude similarity deviation (GMSD) \cite{xue2013gradient})  are used to evaluate the qualities of images synthesized by the proposed method. In the segmentation task, we employ four metrics, detection quality (DQ) \cite{DBLP:journals/corr/abs-1801-00868}, segmentation quality (SQ) \cite{DBLP:journals/corr/abs-1801-00868}, panoptic quality (PQ) \cite{DBLP:journals/corr/abs-1801-00868}, and
aggregated Jaccard index (AJI)  \cite{kumar2017dataset} to evaluate segmentation models trained using different combinations of datasets. 

\textbf{Training process}. The generator input size is 256×256, batch size is 1, and the learning rate is 0.0001. We set epochs to 150.

\begin{figure}
\begin{center}
\begin{subfigure}[b]{0.24\linewidth}
   \vspace{2mm}
    \includegraphics[width=\linewidth]{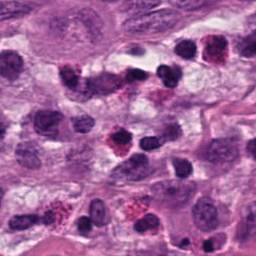}
  \end{subfigure}
  \begin{subfigure}[b]{0.24\linewidth}
    \vspace{2mm}    
    \includegraphics[width=\linewidth]{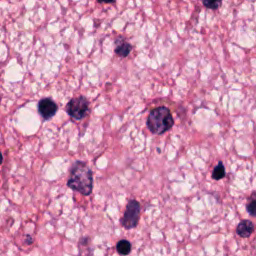}
  \end{subfigure}
  \begin{subfigure}[b]{0.24\linewidth}
    \vspace{2mm}    
    \includegraphics[width=\linewidth]{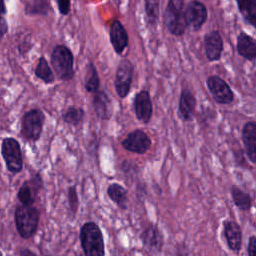}
  \end{subfigure}
  \begin{subfigure}[b]{0.24\linewidth}
    \vspace{1mm}    
    \includegraphics[width=\linewidth]{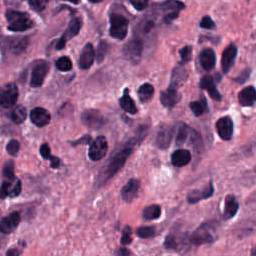}
  \end{subfigure}
   
  \vspace{2mm}
   \vspace{1mm}
  \begin{subfigure}[b]{0.24\linewidth}
    \vspace*{-2.5mm}    
    \includegraphics[width=\linewidth]{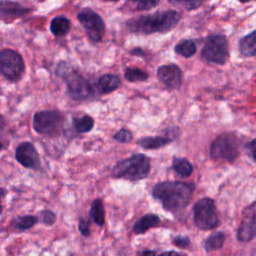}
  \end{subfigure}
  \begin{subfigure}[b]{0.24\linewidth}
    \vspace*{-2.5mm}    
    \includegraphics[width=\linewidth]{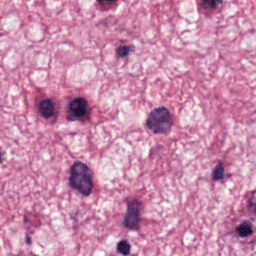}
  \end{subfigure}
  \begin{subfigure}[b]{0.24\linewidth}
    \vspace*{-2.5mm}    
    \includegraphics[width=\linewidth]{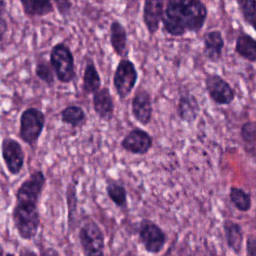}
  \end{subfigure}
  \begin{subfigure}[b]{0.24\linewidth}
    \vspace*{-2.5mm}    
    \includegraphics[width=\linewidth]{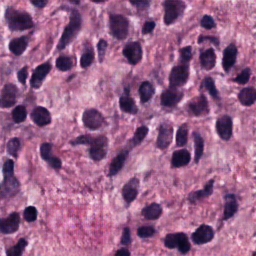}
  \end{subfigure}
\end{center}
    \vspace*{-6mm}
   \caption{First row: Real histopathology images. Second row:
synthesized images from the proposed model. Synthetic images show clear nuclei boundaries, and their colors and textures  are close to the real images.}
\label{fig:long}
\label{fig:onecol}
\end{figure}

\begin{figure}
\begin{center}
  \begin{subfigure}[b]{0.24\linewidth}
    \vspace*{-2.5mm}    
    \includegraphics[width=\linewidth]{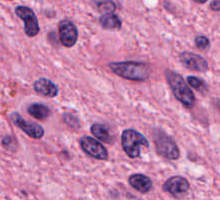}
  \end{subfigure}
  \begin{subfigure}[b]{0.24\linewidth}
    \vspace*{-2.5mm}    
    \includegraphics[width=\linewidth]{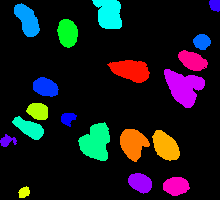}
  \end{subfigure}
  \begin{subfigure}[b]{0.24\linewidth}
    \vspace*{-2.5mm}    
    \includegraphics[width=\linewidth]{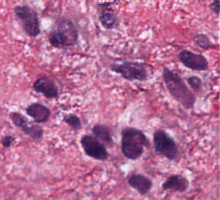}
  \end{subfigure}
  \begin{subfigure}[b]{0.24\linewidth}
    \vspace*{-2.5mm}    
    \includegraphics[width=\linewidth]{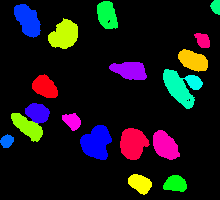}
  \end{subfigure}
\end{center}
    \vspace*{-6mm}
   \caption{Original image patch (left) and generated patch (right) along with segmentation results using pre-trained SegNet on reals.}
\label{fig:long}
\label{fig:onecol}
\end{figure}

\begin{table*}[t]
\small
\begin{center}
\begin{tabular}{l|cccc|cccc}
\hline
Dataset & \multicolumn{4}{c|}{CPM-15\&17} & \multicolumn{4}{c}{TNBC}\\
\hline
Methods & SSIM$\uparrow$ & FSIM$\uparrow$ & GMSD$\downarrow$ & AJI$\uparrow$ & SSIM$\uparrow$ & FSIM$\uparrow$ & GMSD$\downarrow$ & AJI$\uparrow$\\
\hline
pix2pix \cite{isola2017image}   & 0.452 &  0.712 & 0.212 & 0.657 & 0.473 & 0.751 & 0.145 & 0.554\\
Sharp-GAN \cite{butte2022sharp} & 0.569 & 0.731 & 0.174 & 0.671 & 0.504 & 0.755 & 0.139 &  0.568 \\

Sharp-GAN+ & 0.594 & 0.740 & 0.165 & 0.677 & 0.513 & 0.756 & 0.134 & 0.572 \\

Ours & \textbf{0.689} & \textbf{0.757} & \textbf{0.156} & \textbf{0.681} & \textbf{0.522} & \textbf{0.760} &  \textbf{0.131} & \textbf{0.580}\\
\hline  
\end{tabular}
\end{center}
\vspace*{-5mm}
\caption{Image quality assessment on CPM and TNBC datasets}
\end{table*}

\begin{table*}
\small
\begin{center}
\begin{tabular}{l|cccc|cccc}
\hline
Dataset & \multicolumn{4}{c|}{CPM-1517} & \multicolumn{4}{c}{TNBC}\\
\hline
Training Set & DQ & SQ & PQ & AJI & DQ & SQ & PQ & AJI\\
\hline
$\bold{R}$ & 0.760  & 0.764 & 0.584 & 0.613 & 0.648 & 0.731 & 0.475 & 0.486\\
$\bold{R^*}$ & 0.818  & 0.791 & 0.649 & 0.650 & 0.724 & 0.753 & 0.545 & 0.522\\
$\bold{S}$ & 0.753 & 0.784 & 0.594 & 0.592 & 0.838 & 0.797 & 0.669 & 0.654\\
$\bold{S^*}$ & \textbf{0.835} & \textbf{0.798} & \textbf{0.668} & \textbf{0.667}  & \textbf{0.855} & \textbf{0.863} & \textbf{0.691} & \textbf{0.683} \\
\hline  
\end{tabular}
\end{center}
\vspace*{-5mm}
\caption{Segmentation performance of SegNet using different training sets.}
\end{table*}

\subsection{Overall performance}
We compare the proposed approach,  pix2pix GAN \cite{isola2017image} and Sharp-GAN \cite{butte2022sharp} using CPM-15 \& 17 and TNBC. Three image quality metrics, SSIM, FSIM, GMSD are used to evaluate the similarity between real and synthetic images. The segmentation metric AJI is employed to assess the utility of generated images. We train the SegNet \cite{badrinarayanan2017segnet} using real images, and apply it to segment the generated images. The AJI is calculated by comparing the segmentation results and nuclei mask. In Table 1, the Sharp-GAN+ adds skeleton map to the Sharp-Gan. Sharp-GAN+ obtains improved performance in four metrics, which validates effectiveness of using skeleton map. On the CPM-15\&17 dataset, the proposed approach achieves a SSIM of 0.689, a FSIM of 0.757, a GMSD if 0.156, and an AJI of 0.681, and outperforms three other approaches. For example, the proposed approach improves the SSIM by 21$\%$ when compare to the Sharp-GAN.On the TNBC dataset, the proposed method also achieves the best performance.

Figs. 1 and 4 show the sample image patches synthesized by our method. The synthetic images show clear nuclei boundaries and follows colors and textures of real image closely. Fig. 5 shows the segmentation results of a pre-trained SegNet using a real image patch and a synthesized image.

\subsection{Nuclei segmentation using synthesized images}
We evaluate the proposed approach using a downstream task:  nuclei segmentation task. SegNet \cite{badrinarayanan2017segnet} is employed as the baseline network to segment nuclei from histopatology images. A post-processing step using the Watershed algorithm separates different nuclei \cite{naylor2018segmentation} to produce the final segmentation results. Six thousand nucleus-likely polygons are generated \cite{hou2019robust} to create skeleton masks. Four SegNet models are trained using four different training sets. Real image set ($\bold{R}$) contains only real histoathology images from the CPM-15\&17 training set; $\bold{R^*}$ denotes the $\bold{R}$ set augmented using the traditional augmentation techniques, such as rotation, flip, median blur, etc; ($\bold{S}$contains synthetic images only; and $\bold{S^*}$ denotes a combined set of real and synthesized images with traditional image augmentations. As shown in Table 2, compared to the results trained using real data, the results of SegNet trained using 6k synthetic images are significantly better on TNBC dataset, and are competitive in CPM15-17 dataset. The final SegNet is trained using a combination of synthetic and real images with augmentation, and it achieves the state-of-the-art performance on two datasets. Fig. 6 shows an example of segmentation result using a SegNet trained on both real and synthesized images. 

\begin{figure}
\begin{center}
  \begin{subfigure}[b]{0.3\linewidth}
    \vspace*{-2.5mm}    
    \includegraphics[width=\linewidth]{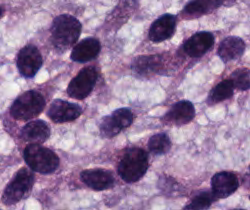}
  \end{subfigure}
  \begin{subfigure}[b]{0.3\linewidth}
    \vspace*{-2.5mm}    
    \includegraphics[width=\linewidth]{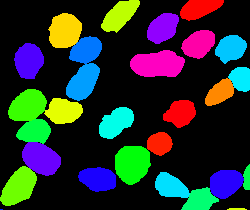}
  \end{subfigure}
  \begin{subfigure}[b]{0.3\linewidth}
    \vspace*{-2.5mm}    
    \includegraphics[width=\linewidth]{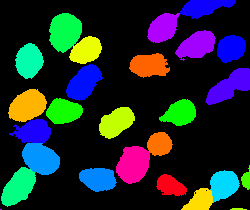}
  \end{subfigure}
\end{center}
    \vspace*{-6mm}
   \caption{An example of nuclei segmentation using SegNet trained on $\bold{S^*}$. From left to right: image patch, image mask, and segmentation result.}
\label{fig:long}
\label{fig:onecol}
\end{figure}

\section{conclusion}
In this paper, we propose an enhanced Sharp-GAN to generate realistic histopathology images. It uses nucleus skeleton map as input to embed nuclei topology and the boundaries between overlapped and touching nuclei. We propose the smoothness loss term to increase the smoothness of contour pixels, and the sharpness loss term to enhance the contrast between contour and non-contour pixels. Extensive experiments on two public datasets demonstrate that the proposed approach achieves the state-of-the-art performance in synthesizing realistic histopathology images. By integrating the synthesized images from the proposed approach into training, ther performance of a segmentation model could be significantly improved.

{\small
\bibliographystyle{unsrt}
\bibliography{refs.bib}
}
\end{document}